\documentclass{emulateapj}

\usepackage{calc}
\newcommand{\grb}{gamma ray burst}
\newcommand{\asa}{ASASSN-15lh}
\newcommand{\snkl}{SN~2011kl}

\newcommand{\Ni}{{$^{56}$Ni}\,}
\newcommand{\Nimass}{{$M(^{56}\mathrm{Ni}$)}\,}

\shorttitle{SLSNe \snkl\ and \asa}
\shortauthors{Bersten, Benvenuto, Orellana and Nomoto}
\begin{document}
\title{The Unusual Super-Luminous Supernovae \snkl\ and \asa}
\author{%
        Melina C. Bersten\altaffilmark{1,2,6},
        Omar G. Benvenuto\altaffilmark{1,2,3}, 
        Mariana Orellana\altaffilmark{4, 5} 
        and 
        Ken'ichi Nomoto\altaffilmark{6,7}\\
}

\altaffiltext{1}{Facultad de Ciencias Astron\'omicas y Geof\'{\i}sicas,
Universidad Nacional de La Plata,  
Paseo del Bosque S/N, B1900FWA La Plata, Argentina}

\altaffiltext{2}{Instituto de Astrof\'isica de La Plata (IALP),
  CONICET, Argentina}  

\altaffiltext{3}{Member of the Carrera del 
Investigador Cient\'{\i}fico de la Comisi\'on de Investigaciones
Cient\'{\i}ficas de la Provincia de Buenos Aires (CIC), Argentina}

\altaffiltext{4}{Sede Andina, Universidad Nacional de R\'{\i}o Negro,
  Mitre 630 (8400) Bariloche, Argentina} 
  
\altaffiltext{5}{Member of the Carrera del 
Investigador Cient\'{\i}fico y Tecnol\'ogico del CONICET, Argentina}

\altaffiltext{6}{Kavli Institute for the Physics and Mathematics of
the Universe, The University of Tokyo Institutes for Advanced Study (UTIAS), 
The University of Tokyo, 5-1-5 Kashiwanoha, Kashiwa, Chiba 277-8583, Japan}

\altaffiltext{7}{Hamamatsu Professor}

\setcounter{footnote}{7}
 \email{mbersten@fcaglp.unlp.edu.ar}  
\submitted{Submitted to ApJL on November 12, 2015. Accepted on December 27, 2015}
\begin{abstract}
\noindent Two recently discovered very luminous supernovae (SNe)
present stimulating cases to explore the extents of the available
theoretical models. \snkl\ represents the first detection of a
supernova explosion associated with an ultra-long  
duration \grb. \asa\ was even claimed as the most
luminous SN ever discovered, challenging the scenarios so far proposed
for stellar explosions. Here we use our radiation hydrodynamics code in
order to simulate magnetar powered SNe. To avoid explicitly assuming 
neutron star properties we adopt the magnetar
luminosity and spin-down timescale as free parameters of the model. We
find that the light curve (LC) of \snkl\ is consistent with a
magnetar power source, as previously proposed, but we note that some amount of
$^{56}$Ni ($\gtrsim 0.08 M_\odot$) is necessary to explain the low contrast
between the LC peak and tail. For the case of \asa\ we find physically
plausible magnetar parameters that 
reproduce the overall shape of the LC provided the progenitor
mass is relatively large (an ejecta mass of $\approx\,$6 $M_\odot$).
The ejecta hydrodynamics of this event is dominated by the magnetar
input, while the effect is more moderate for \snkl.
We conclude that a magnetar model may be used for the
interpretation of these events and that the hydrodynamical modeling is
necessary to derive the properties of powerful magnetars and
their progenitors. 
\end{abstract}

\keywords{ 
stars: evolution ---  
supernovae: general --- 
supernovae: individual (SN~2011kl) ---
supernovae: individual (ASASSN-15lh) ---
gamma-ray burst: individual (GRB 111209A)
}

\noindent\section{INTRODUCTION} \label{sec:intro}

Superluminous supernovae (SLSNe) have been discovered almost a
decade ago \citep{Q07,S07}. They show a factor 10 to 100 times
brighter than normal core-collapse supernovae (SNe), often well
above $-22$ 
absolute magnitude. It is believed that they are the explosion of
massive stars which usually are found in low luminosity, star forming
dwarf galaxies 
\citep{N11,L14,L15}. But the physical origin of the extreme luminosity
emitted remains speculative \citep[see, e.g.,][for a review]{G12}. The
high luminosity of SLSNe makes them ideal to get information from the
early universe and to explore the capability to use them as
cosmological standard candles at much further distance than normal
SNe \citep{Q11,IS14}.
  
In order to be radioactively powered, as normal SNe, SLSNe
would require a too large nickel mass except for pair instability
  SNe and hypernovae \citep[see, e.g.,][]{Moriya10}, 
so competing, so competing
alternatives have been proposed. One possible mechanism invoked to
explain SLSNe is the energy  
injection by an accreting black hole that launches relativistic jets, or
fallback scenario \citep[e.g.,][and
  references therein]{DK13}. The interaction of the SN ejecta
with dense circumstellar material (CSM) is another proposed
mechanism \citep[see e.g.][]{S15}. Another popular
explanation is that a magnetar is formed by the collapse of a massive
star. The 
magnetar is a strongly-magnetized, rapidly-rotating neutron star that
loses rotational energy via magnetic dipole radiation.   
Although some progress in the area were recently reported
\citep[see e.g.][]{TH15}, other important aspects of
the scenario are still unclear.  

In this work we study two peculiar SLSNe of Type Ic (lacking
hydrogen),  \snkl\, and for \asa.
\snkl\, has been associated with the ultra-long-duration \grb ~GRB
111209A, at a redshift $z$ of 0.677  
\citep{G15}. Its light curve (LC) was significantly over-luminous
compared to other GRB-associated SNe, suggesting for the first time a
link between SN-GRB and SLSNe-ultra long GRB (ULGRB). The precise
explosion time estimation makes this SN ideal for numerical modeling.

More recently, \asa ~at $z$ = 0.2326 was discovered by the All-Sky
Automated Survey for SuperNovae \citep{D15}. It showed an hydrogen-poor
spectrum, and the maximum luminosity was $\sim2.2\times 10^{45}$
erg s$^{-1}$, i.e. more luminous than any previously known SN. Contrary to
SN 2011kl, the explosion date of \asa ~is 
unknown, and data before the luminous peak were less reliable than
later. The optical emission was continuously detected for more than
100 days after the explosion. The spectrum lacked the broad
H$\alpha$ emission feature that would have evidenced interaction between
the supernova and an H-rich circumstellar environment \citep{MJ15}. 

Magnetar models have been proposed for the two SLSNe of this work
\citep{G15,M15}, even for the extremely luminous \asa. However,
the analysis was based on simplistic assumptions
that neglected the dynamic effects on the ejecta. Here we shall study 
the magnetar scenario using hydrodynamic calculations which
incorporate the dynamical effect and considering the limit imposed by
the neutron star (NS) matter equation of state (EOS).   
 
\section{MAGNETAR MODELS} \label{sec:numeric_CLs}

We include an extra source of energy due to a rapidly rotating
and strongly magnetized NS (or ``magnetar'') in our one-dimensional
LTE radiation hydrodynamic code \citep{BBH11}. A spherical young
magnetar releases its rotational energy at a well known rate described
by the radiating magnetic dipole
\citep[e.g. ][]{1983bhwd.book.....S}. 
We assume that this energy is fully deposited and thermalized in the
innermost layers of our pre-supernova model. This assumption as well
as a large inclination $i= 45^\circ$ is 
the same used in previous numerical works of magnetar powered SN light
curves \citep{W10,KB10}. 
Powerful enough magnetars may force the
envelope to expand at velocities comparable to the speed of light (see
below, particularly Fig.~\ref{fig:veloasa}). Therefore, we have
modified our code to properly include relativistic velocities. A
detailed description of our treatment of relativistic radiating
hydrodynamics will be presented in a forthcoming paper. 

Our hydrodynamical calculations simulate the explosion of an
evolved star, followed consistently until core collapse
condition. The pre-SN He star models of different masses used in this
paper were calculated by \citet{NH88}. The explosion itself is
simulated as usual 
by the injection of a certain energy (a few $\times 10^{51}$~erg) at the
innermost layers of our pre-SN models. In addition to the
explosion energy, the rotational energy lost by newly born magnetar
is included. 

Once the NS momentum of inertia and radius $R$ are fixed, this energy
source essentially depends on two parameters, the strength of the
dipole magnetic field\footnote {Considered fixed during the
evolution}, $B$, and the initial rotational period, $P_0$ (see,
e.g., equations 1 and 2 of \citet{KB10}).  It is equally possible to
use the spin-down timescale ($t_{\rm p}$) and magnetar
energy loss rate ($L_{\rm p}$) as the free parameters to be determined
by fitting the observed light curve (LC). Therefore the magnetar
luminosity can be written as 
\begin{equation}
L= L_{p} \bigg(1+\frac{t}{t_{p}}\bigg)^{-2},
\end{equation}
with
\begin{equation}
L_{p}= \frac{4 \pi^4 R^6}{3 c^3} \frac{B^2}{P^4},
\end{equation}
and
\begin{equation}
t_{\rm p}= \frac{3 c^3 I_{NS}}{\pi^2 R^6} \frac{P^2}{B^2}.
\end{equation}

\noindent The values of $B$ and $P_0$ can be found later from $t_{\rm p}$,
$L_{\rm p}$, assuming an 
NS mass. Then,  the radius and momentum of inertia are
found for a given EOS. In this way, one avoids to assume from
beforehand explicit 
values of the NSs properties. This method opens the possibility to
accommodate a variety of NS structures, which is relevant because of our 
poor knowledge of the EOS at these extreme conditions.    

Along the calculation, if the photosphere recedes deep enough so that part of the magnetar energy is directly deposited outside the 
  photosphere, we add this energy to the bolometric luminosity. The 
  same treatment is used for $^{56}$Ni deposition
  \citep{S91,BBH11}. We implicitly assume that the magnetar
  (or the $^{56}$Ni decay) produces hard photons
  than can be trapped even if the ejecta is
  optically thin to optical photons. During these epochs, usually
  after about 60 days (although this depends on the
  progenitor mass and energy), the emitted luminosity is almost the
  same as the magnetar luminosity.  

\section{Results}

\subsection{Models for \snkl}

\citet{G15} (hereafter G15), recently analyzed the LC and spectrum of
\snkl\,. A low ejecta mass $\approx$ 2.4 M$_\odot$ and high explosion
energy $\approx$ 5.5 $\times 10^{51}$ erg were obtained from their
analysis. The radiative output of \snkl\, locates it in between the
GRB-SNe and SLSNe. A large $^{56}$Ni mass $\approx$ 1 M$_\odot$ was
found necessary in order to reproduce the high observed
luminosity. This large amount of radioactive material seems to be
neither compatible with the spectrum  
properties nor with the low mass ejecta, as pointed by G15. A
magnetar source was then proposed by the authors to explain the LC.  

We used the code describe in ~\S\ref{sec:numeric_CLs} to study
\snkl. One advantage of the hydrodynamics calculations is that the
explosion time ($t_{\rm exp}$) is fully determined from the assumed physical
parameters. The GRB detection of \snkl\, establishes a tight value of $t_{\rm
  exp}$\footnote {Here we assume that the GRB precursor
  indicated the explosion itself. But see \citet{VS99} for other possible
  scenarios.} then, it is an ideal target of study. Motivated by the
values proposed by G15 we assumed a pre-SN He star of 4 M$_\odot$ 
with a cut mass (M$_{\rm cut}$) of 1.5 M$_\odot$ corresponding
to an ejected mass ($M_{\rm ej}$) of 2.5 M$_\odot$ and an explosion energy of 
5.5$\times 10^{51}$ erg. While the LC of \snkl\, is unlikely to be powered only
by \Ni some amount of radioactive material is usually expected to be
produced during the explosion. We assume a  \Ni\, mass of 0.2
M$_\odot$ consistent with radioactive yield of energetic SN
\citep{N13}.   

\begin{figure} [h]
\resizebox{\hsize}{!}{\includegraphics[angle=-90]{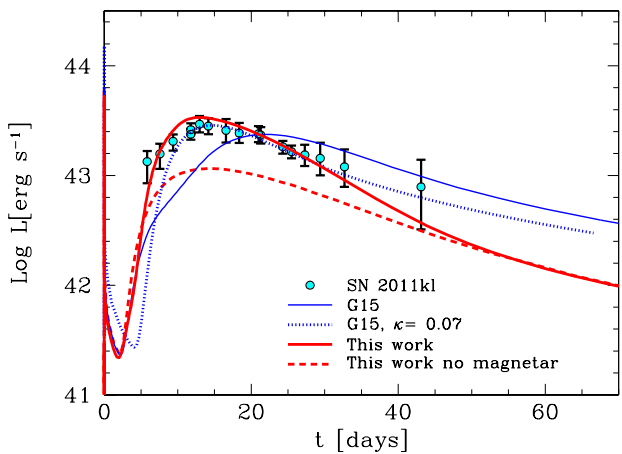}}
\caption{Bolometric LC of \snkl\, compared with several LC
  models. Thick red solid line shows our preferred model with
  $P_0= 3.5$ ms, and magnetic field  $B=1.95\times 10^{15}$~G (see text
  for more details). A model with $P_0=12.2$ ms and $B=7.5\times
  10^{14}$~G as suggested by G15 computed with our hydrodynamical
    modeling and our opacity prescription is
  shown in thin blue line and in dotted line for $\kappa$= 0.07 cm$^2$
  g$^{-1}$. A $^{56}$Ni power model with a $^{56}$Ni mass of 1
  M$_\odot$ is shown with dashed line. 
\label{fig:CL11kl}}
\end{figure}

We extensively explored the magnetar parameter space and
found a reasonable agreement with the data for $P_0=3.5$ ms and magnetic  
field $B=1.95\times10^{15}$~G. Figure~\ref{fig:CL11kl} shows this model
with thick solid line. Our values lie out of the ranges given by G15
whose analysis gave $P_0=12.2\pm 0.3$~ms and  $B=7.5 \pm 1.5\times
10^{14}$~G. The thin line of Figure~\ref{fig:CL11kl} show the
  result of our hydrodynamical calculations assuming the same
  parameters as in G15. We found a poor match to the
data as opposed to G15 (see their Fig.~2). We note that we have used OPAL
opacity tables with an opacity floor of 0.2  cm$^2$ g$^{-1}$
corresponding to the electron scattering opacity for hydrogen free
material. This value is usually assumed as gray opacity in LC magnetar
models in the literature \citep{KB10,I13}. However, with a lower and
constant value of $\kappa=$ 0.07 cm$^2$ g$^{-1}$, presumably used in
G15, we 
recovered a reasonable fit to the data for the magnetar parameters of
G15 (see dotted line of Figure~\ref{fig:CL11kl}). 

In addition, we present our results for a model powered only by
$^{56}$Ni. A model 
with \Nimass $=$ 1 M$_\odot$ is shown with a dashed line in
Figure~\ref{fig:CL11kl}. Even with this 
large amount of $^{56}$Ni we could not reach the high luminosity
needed to explain \snkl\, using our opacity prescription. While
an non-standard source of energy is necessary to explain the peak
luminosity, some amount of $^{56}$Ni
($\gtrsim$ 0.08 M$_\odot$) is also needed to explain the tail
luminosity. Magnetar models without nickel produce a larger contrast
between peak and tail and thus a poor fit to the data.
 
\begin{figure} [hb]
\resizebox{\hsize}{!}{\includegraphics[angle=-90]{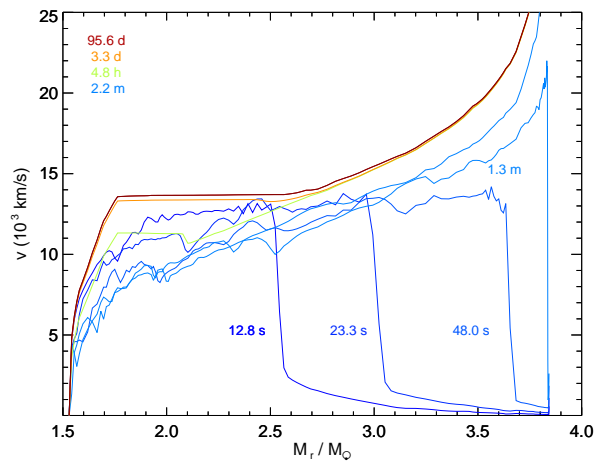}}
\caption{Velocity profile for the explosion that reproduce the LC of
  \snkl. The interior dynamics is affected in comparison to a standard
  explosion without any magnetar. A day after explosion the bulk of
  the stellar mass expands at $\approx 13\times 10^3$ km~s$^{-1}$, or higher,
  whereas the outermost layers (not shown in the figure) reach
  velocities even higher (up to $0.21\; c$). 
\label{fig:velo11kl}}\end{figure}

Figure~\ref{fig:velo11kl} shows the velocity profile at different
times since the explosion for our preferred model (solid thick line of
Fig.~\ref{fig:CL11kl}). Initially, the shock wave propagates as in a
standard explosion. Later the
dynamic effect is noticeable although not as extreme as the case of
\asa\, (see below). The extra heating source due to the magnetar swell the inner
zones and produces larger velocities and a flat profile that raise
steeply only at the outermost layers. The high ejecta velocity is
consistent with the analysis by G15 who inferred that a large
value of $v_{\rm ph}$ would explain the rather featureless spectrum by
Doppler line blending. We remark that here the opacity
  prescription is probably  
  responsible for the differences found between our calculation and
  the analytic models presented by G15. The hydrodynamic effects of
  the magnetar on the luminosity are not as important as in the case
  of \asa, although as shown in Figure~\ref{fig:velo11kl}, it is
  evident in the velocity evolution.

\subsection{Models for \asa}

\citet{D15} reported the discovery of the brightest known SN, \asa\,
with a peak luminosity of $\approx$ 2 $\times 10^{45}$ erg s$^{-1}$,
almost two magnitudes brighter than the average observed SLSNe, and with
total radiated energy  of $\approx$ $10^{52}$ erg. The spectra
presented some similarities to SLSNe-Ic. Magnetar models seem to be
the most plausible explanation for this class of SLSNe. However,
\citet{D15} emphasized the difficulty to explain this object in the
magnetar context. The short initial period $\approx$ 1 ms required to
match the LC properties could be in contradiction with the maximum
rotational energy ($E_{\rm max}$) available that in turn is limited by
the emission of gravitational waves. However, \citet{M15} lately
suggested that $E_{\rm max}$ could be up to one order of magnitude
larger than previous estimations, depending on the NS
properties, but avoiding the problem raised by \citet{D15}. 

\begin{figure}[h] 
\resizebox{\hsize}{!}{\includegraphics[angle=-90]{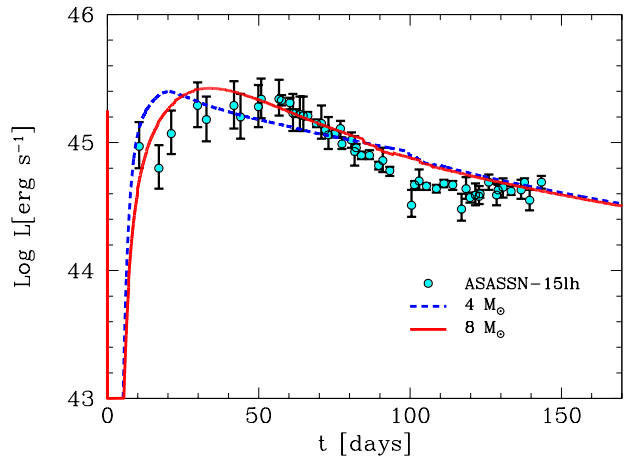}} 
\caption{Observed bolometric LC of ASASSN-15lh (dots; \citealt{D15})
  compared with models of 4 M$_\odot$ (dashed line) and 8 M$_\odot$
  (solid line) pre-SN mass for magnetar parameters of $L_{\rm p}=9
  \times 10^{45}$ and $t_{\rm p}= 40$ days and $t_{\rm exp}=$ JD 2457143.     
\label{fig:CL15lh}}\end{figure}

We calculated a set of magnetar models to analyze the LC of
\asa\,. Here we use $L_p$ and $t_{\rm p}$ as free parameters for the
fitting (see \S~\ref{sec:numeric_CLs}). In this way, we avoid
including explicitly the properties of the 
NS. Once the LC fit is admissible, we derived $B$ and $P_0$ as a
function of the magnetar mass (see bellow). Figure~\ref{fig:CL15lh}
shows our results for a pre-SN He star of 4 M$_\odot$ with  
M$_{\rm  cut}$= 1.5 M$_\odot$  and M$_{\rm ej}=$ 2.5 M$_\odot$ (thin
red line) and for a more massive model of He star 8 M$_\odot$ with
M$_{\rm cut}$= 2 M$_\odot$ and M$_{\rm  ej}=$ 6 M$_\odot$ (thick blue line). 
We assumed an initial explosion energy of 5.5$\times 10^{51}$ erg
although this has a minor effect on the results. Also, we 
assumed that the SN exploded a week before the discovery. 

The magnetar parameters are in this case $L_{\rm p}=9 \times10^{45}$
erg s$^{-1}$ and $t_{\rm p}= 40$ days. For the less massive model we
could not  
reproduce the overall LC shape. A more massive pre-SN model with 8
$M_\odot$ or more is therefore needed to reproduce the temporal width
of the LC.  
At $\sim$ 100 days, there is a slight change on the LC slope in
coincidence with the moment that the photosphere reaches the inner
regions where the magnetar energy is directly deposited. The
observable effect of this fact should potentially modify the spectral
energy distribution, and deserve further investigation. 

\begin{figure} \begin{center}
\includegraphics[scale=.3,angle=0]{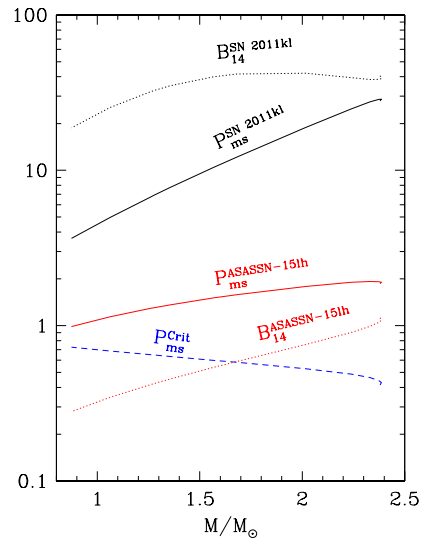}
\caption{Magnetic fields (in units of $10^{14}$~Gauss, represented
  with dotted lines) and initial spin periods (in milliseconds,
  denoted with solid lines) of the magnetar as a  function of its
  mass. Black lines correspond to the parameters for \snkl\, whereas
  red lines denote the case of ASSASN-15lh. For comparison, the
  critical spin period is given with a dashed line. Notice that for
  both cases the NS can rotate at the required rate.  
\label{fig:valores}} \end{center} 
\end{figure}

Figure~\ref{fig:valores} shows the values of $B$ and $P_0$ as a function
of the NS mass derived from  $L_{p}$ and $t_{\rm p}$. For the
structure of the magnetar we have assumed the data presented in Table
12 of \citet{A77}. This corresponds to the structure of a NS assuming the
nuclear matter equation of state of \citet{B75} that gives a mass
radius relation similar to those currently favored by recent
observations (see Fig.~11b of \citealt{L12}). We have neglected
rotational and magnetic effects on the NS structure. For comparison,
we included the curve for the case of \snkl\, which represents
the location of other possible solutions (i.e. 
 the degeneracy of the parameters $B$ and $P_0$). 
Physically possible solutions correspond to rotation periods
larger than the critical breakup value.  
For the complete NSs mass range, we found solutions that fulfill this
condition.    
For the pre-SN model a specific value of M$_{\rm cut}$ was assumed,
but we have corroborated that changing this value in the range shown in
Figure~\ref{fig:valores} produces a minor effect on the LC model.   

We conclude that initial periods ranging $\sim 1 - 2$ ms and magnetic
fields of $\sim (0.3 - 1)\times 10^{14}$~G  related by the curve of
Figure~\ref{fig:valores} provide a reasonable fit to the LC. These
magnetar values are in good agreement with those proposed by
\cite{M15} although they assumed a lower ejecta mass of 3
M$_\odot$. However, we could not reproduce the data with such low mass
progenitor.  

Here we should note that the uncertainty regarding $t_{\rm exp}$
could modify the exact value of the parameters. In any case,
the scope of this analysis is to show that the magnetar scenario is
plausible for this object and not to provide definitive values of the
physical parameters.  

\begin{figure}
\resizebox{\hsize}{!}{\includegraphics[angle=-90]{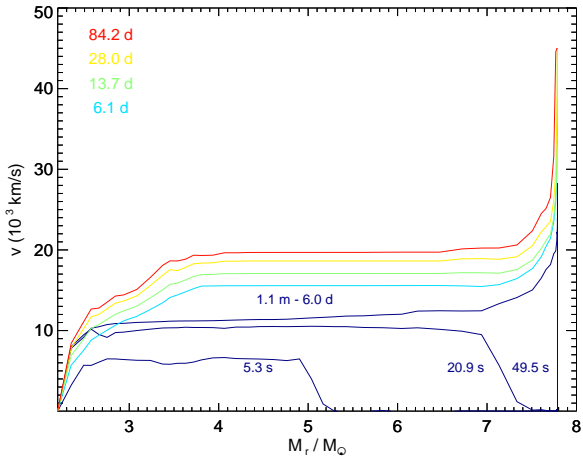}}   
\caption{Velocity profile for \asa\, magnetar model with $L_{\rm p}=9
  \times 10^{45}$ and $t_{\rm p}= 40$ days. This extreme case inflates
  the external radial zones up to $0.15\, c$.  
\label{fig:veloasa}} 
\end{figure}

Figure~\ref{fig:veloasa} shows the velocity profiles for our preferred
model of 8 M$_\odot$. In this case the magnetar is extremely powerful
and the dynamical effect is more noticeable than for the case of
\snkl. The energy permanently supplied  by the magnetar impulses all
the envelope at high 
velocities and particularly the outermost layers. As result, a few
days after the explosion most of the material moves at constant
velocity, which increases with time due to
the permanent injection of magnetar energy.  
This dynamic behavior should have a clear effect on the
spectrum. Broad line features at 4100 and 4400 \AA\, were observed in
the optical spectrum of \asa\, between 13 and 20 days after maximum
implying very high velocities of $\approx$ 20.000 km s$^{-1}$
\citep{M15}. This timing and velocity are fully consistent with the
model shown in Figure~\ref{fig:veloasa}. We note that 
the photosphere at this epoch is located in the flat part of the
velocity profile.   

\section{Discussion and Conclusions}
We were able to reproduce the LC of \snkl\, and \asa\, in the context
of magnetar-powered models with physically allowed parameters
(the NS rotating below breakup point).   

By adopting the magnetar luminosity and spin-down timescale as free
parameters we could separate the LC fit from any assumption on the NS
structure. The usual parameters ($B$ and $P_0$) could
be recovered afterwards by assuming the NS configuration for a given
EOS. We have shown that this leads to a degeneracy between $B$ and
$P_0$ which in turn depend on the properties of the nuclear matter EOS
which is poorly known.

For \snkl\, we found $L_{\rm p}\approx 1.2 \times 10^{45}$ erg s$^{-1}$
and $t_{\rm p}\sim 15$ days. A family of values of $P_0$ and $B$ 
were derived that lie outside the ranges proposed by G15.
We ascribe the differences to the opacity values adopted and not
  to the hydrodynamic effects. 
We note that in addition to the extra source of energy due to
a magnetar, some non-negligible amount of $^{56}$Ni 
($\gtrsim$ 0.08 M$_\odot$) was also necessary in our calculation
to produce the LC peak and tail.

Regarding the extreme luminosity of \asa\, the overall shape of the LC
was reproduced for $L_{\rm p}\approx 9 \times 10^{45}$ erg s$^{-1}$
and $t_{\rm p}\sim 40$ days. The resulting ranges of $B$ and $P_0$ were
compatible with those proposed by \cite{M15}. However,
we note that our numerical models require a
massive progenitor with $M_{\rm ej} \approx$ 6
$M_\odot$, i.e. a factor of 2 larger than the value adopted by
\cite{M15}. Translating $M_{\rm ej}$ into a zero-age
main-sequence mass, $M_{\rm  ZAMS}$, of the progenitor involves
several uncertainties. It is particularly important whether the 
He features can be hidden in the early spectra of SLSNe
\citep{H12,D12}. Assuming that this is the case, the pre-SN star
could be a He star of 8 $M_\odot$, which corresponds to 
$M_{\rm ZAMS}\sim$ 25 $M_\odot$ \citep[see][for a relation between pre-SN mass
and $M_{\rm ZAMS}$]{T09}. On the other hand, if no He is present, then
the pre-SN star could be a C+O star of 8 M$_\odot$, which corresponds to a
$M_{\rm  ZAMS} \sim$ 30 $M_\odot$. In both cases, the
progenitor mass is close to the boundary mass between  
BH and NS formation \citep{N13}. We emphasize that hydrodynamical
modeling of the LC can provide better constraints on the highly
uncertain progenitor masses of magnetars than the analytic
prescription. 

Our treatment of the SN evolution illustrates the importance of the
dynamical effects on the ejecta, especially in cases of powerful
magnetars. The homologous expansion, usually assumed in SN
studies, can be broken because of the additional energy source. 
This could have an important effect on the line formation and the
photospheric velocity evolution. For \asa\, we found a total energy
release by the magnetar of $E\sim 3 \times 10^{52}$ erg, which is one
order of magnitude larger than the initial explosion energy. For the
more moderate \snkl\, we obtained $E \sim 1.6 \times 10^{51}$ erg. In
general, the dynamical effects on the expansion of the ejecta become
significant when the magnetar energy is comparable with the explosion
energy. 


\acknowledgments 
 We are grateful to Sergei Blinnikov and Gast\'on Folatelli for the
 helpful conversations and to Jose Prieto for sending the data of \asa.   
This research is supported by the World Premier International Research
Center Initiative (WPI Initiative), MEXT, Japan and the Grant-in-Aid
for Scientific Research of the JSPS (23224004, 26400222), Japan.
M.O. would like to thank Kavli IPMU for the support for her visit to
Kavli IPMU. M.O. also acknowledges UNRN partial support through grant 2014PI 40B364.
 

\clearpage
\end{document}